\documentclass[useAMS]{mn2e}

\usepackage{epsfig}
\usepackage{graphicx}

\def\lsim{\mathrel{\rlap{\lower4pt\hbox{\hskip1pt$\sim$}}
    \raise1pt\hbox{$<$}}}                % less than or approx. symbol
\def\gsim{\mathrel{\rlap{\lower4pt\hbox{\hskip1pt$\sim$}}
    \raise1pt\hbox{$>$}}}                % greater than or approx. symbol

\title[The inner structure of massive elliptical galaxies]{The inner structure of very massive elliptical galaxies: implications for the inside-out formation mechanism of $z\sim 2$ galaxies}

   \author[O.Tiret et al.]{O. Tiret$^1$, P. Salucci$^1$, M. Bernardi$^2$, C. Maraston$^3$, J. Pforr$^3$   \\
   $^1$ SISSA, via Bonomea, 265, 34136 Trieste, Italy\\
                    $^2$ Department of Physics and Astronomy, University of Pennsylvania, 209 South 33rd Str., Philadelphia, PA 19104, USA\\
                     $^3$ Institute of Cosmology and Gravitation, Dennis Sciama Building, University of Portsmouth, Burnaby Road, Portsmouth, PO1 3FX, UK}

\begin{document}

\date{Accepted xxxx. Received xxxx; in original form xxxx}

\pagerange{\pageref{firstpage}--\pageref{lastpage}} \pubyear{xxxx}

\maketitle

\label{firstpage}

\begin{abstract}
We analyze a sample of 23 supermassive elliptical galaxies (central velocity dispersion larger than 330 km s$^{-1}$), drawn from the SDSS.  For each object, we estimate the dynamical mass from the light profile and central velocity dispersion, and compare it with the stellar mass derived from stellar population models. We show that these galaxies are dominated by luminous matter within the radius for which the  velocity dispersion is measured.  We find that the sizes and stellar masses are tightly correlated, with $R_e\propto M_*^{1.1}$, making the mean density within the de Vaucouleurs radius a steeply declining function of $M_*$:  $\rho_e\propto M_*^{-2.2}$.  These scalings are easily derived from the virial theorem if one recalls that this sample has essentially fixed (but large) $\sigma_0$.  In contrast, the mean density within 1 kpc is almost independent of $M_*$, at a value that is in good agreement with recent studies of $z\sim 2$ galaxies.  The fact that the mass within 1 kpc has remained approximately unchanged suggests assembly histories that were dominated by minor mergers -- but we discuss why this is not the unique way to achieve this.  Moreover, the total stellar mass of the objects in our sample is typically a factor of $\sim 5$ larger than that in the high redshift ($z \sim 2$) sample, an amount which seems difficult to achieve. 
If our galaxies are the evolved objects of the recent high redshift studies, then we suggest that major mergers were required at $z\gsim 1.5$, and that minor 
mergers become the dominant growth mechanism for massive galaxies at $z\lsim 1.5$. 
\end{abstract}


\section{Introduction}

The study of luminous and dark matter in elliptical galaxies is crucial
to understanding the formation of massive galaxies in our Universe.
In hierarchical models, ellipticals are the result of interactions and
mergers of spiral galaxies (e.g. Blumenthal et al. 1984).  This is in
contrast to a scenario in which they form from a monolithic collapse
(e.g. Eggen et al. 1962; Granato et al. 2004).
The most massive elliptical galaxies, with baryonic masses
$M>10^{11}M_{\odot}$, are challenging because they should
have formed in the very early universe and at the same time undergone a
large deal of merging.

There is now growing evidence that massive galaxies ($M_* \sim 10^{11}$~M$_\odot$) did
exist at $z\sim 2$. Some work suggests that they were much smaller and denser
than their local counterparts of the same stellar mass
(e.g. Trujillo et al. 2006; van Dokkum et
al. 2008; Cimatti et al. 2008; Saracco et al. 2009) and that
similar compact galaxies to those observed at high-redshift
do not exist in the local universe (e.g. Trujillo et al. 2009).
These results raised the question of what process or processes have acted
to increase the sizes of these objects to make them consistent with the larger sizes we see
at late times (e.g. van Dokkum et al. 2008; Fan et al. 2008).

Bezanson et al. (2009) showed that the stellar density
within the central 1~kpc of ellipticals at $z\sim 2.3$ is similar
to that for nearby ellipticals (they differ by only a factor of $\sim 2$,
compared to a difference of a factor of $\sim 100$ if the comparison is done
within the half-light radius).  This suggests an
inside-out, hierarchical growth scenario dominated by dry minor mergers
which add mass primarily to the outer regions (e.g. Loeb \& Peebles 2003;
Bournaud et al. 2007; Naab et al. 2007; Hopkins et al. 2009).

However, the evidence for small sizes at high-redshift and the lack of such
objects at low-redshift is not uncontested.  For example,
Mancini et al. (2010) have argued that neglect of low surface brightness
features will bias $r_e$ to small values. Their analysis shows that some of
the objects at $z\sim 1.5$ are not small for their $M_*$ compared to $z=0$
objects. A similar result was recently presented by Onodera et al. (2010)
who found a $z=1.82$ analog of local ultra-massive elliptical galaxies.
Recently, Saracco et al. (2010) found from a complete sample of 34 early-type galaxies
at $0.9<z<1.92$  that 21 of these are similar to the local ones even though they are co-eval
with more compact early-type galaxies.  In addition, Valentinuzzi et al. (2010) found that about
25\% of the objects with $M_* > 3\times 10^{10}$~M$_\odot$ in local clusters are superdense
(i.e. they have sizes like those observed out to z $\sim 2$).
However, they found that there is strong evidence for a large evolution in radius
for the most massive galaxies, i.e. BCGs ($M_* > 4\times 10^{11}$~M$_\odot$).

Indeed, evolution in the properties of local BCGs was detected by Bernardi (2009), who showed that the sizes
and velocity dispersions of BCGs (and of massive early-type galaxies) in the local Universe
($z < 0.3$) are still evolving. This work suggests that minor dry mergers
dominate the assembling of BCGs at lower redshifts since the observed size 
evolution is more rapid than expected by major dry mergers once one accounts
for the small changes in the observed luminosity/mass functions
($\sim 50$\% since $z < 1$; e.g. Wake et al. 2006; Brown et al. 2007; Cool et al. 2008).
Minor dry mergers are better able to reconcile the observations of size evolution with little
mass evolution. In addition, Bernardi (2009) also claims that minor mergers
can also help to reconcile the substantial growth ($\sim$ a factor of 2) predicted for the dark matter
halos since $z \sim 1$ (e.g. Sheth \& Tormen 1999) with the little
stellar mass evolution in the central galaxies (i.e. $\sim 50$\%): indeed,
the fractional mass growth of BCGs need not be the same as that of their
host clusters -- some of the added stellar mass must make the intercluster
light.

 However, while an inside-out, hierarchical growth scenario dominated by minor 
dry mergers can describe the assembling of BCGs at low redshift, the 
recent analysis of Bernardi et al. (2010b) suggests that some of the features 
observed in the scaling relations of massive early-type galaxies at 
$M_* > 2\times 10^{11}M_\odot$ (e.g. the upwards curvature in the color$-M_*$ 
relation, the decrease in the mean axis ratio and color gradients and the fact
that most scaling relations with $\sigma$ are well-described by a single power 
law) can only be explained by an assembly history dominated by major dry 
mergers above this mass.

The main goal of this paper is to use both visible stellar masses and dynamical stellar masses, for 23 supermassive elliptical galaxies 
identified by Bernardi et al. (2006, 2008), to investigate these issues in the light of previous work.

The paper is organized as follows.  We describe the observables in 
Section~\ref{obs}, our procedure for estimating the total dynamical 
masses, and the fraction which is in stars, in Section~\ref{jeans}, 
a comparison of these dynamical mass estimates with those from 
stellar population models in Section~\ref{ssp}, 
and scaling relations between size and mass in Section~\ref{reM}.  
A final section discusses our findings:  we argue that it is not 
clear that minor mergers since $z\sim 2$ can have been the dominant 
formation mechanism of these massive galaxies.  

When necessary, we assume a flat background cosmology that is dominated 
by a cosmological constant at the present time:  $\Omega_0=0.3$, 
$\Lambda_0=1-\Omega_0$, and we set Hubble's constant to 
$H_0 = 70$~km~s$^{-1}$Mpc$^{-1}$.  

\begin{table}
\caption{Properties of the 23 supermassive elliptical galaxies of our sample. 
Galaxies are identified using the same index, column(1), as in 
Hyde et al. (2008). 
Column (2) gives the velocity dispersion within the 3 arcsec SDSS fiber; 
Column (3) the projected half-light radius from fitting to a de Vaucouleurs profile in the $i$-band;
Column (4-5) the projected half-light radius and Sersic index from fitting to a Sersic profile in the $i$-band; and 
Column (6), the physical scale on which the velocity dispersion was measured (corresponding to 3 arcsec).
}
\label{tab:data}
\begin{center}
\begin{tabular}{rrrrrrr}
 
 \hline
 \#& $\sigma_0 $     & r$_e^{DeV}$ &  r$_e^{S}$&$n^S$ & r$_{ap}$ \\
   & (km~s$^{-1}$)   & (kpc) & (kpc)& &(kpc)\\
 \hline
1 & 339 &  8.5  & 19.031 & 5.698  &7.5 \\
2 & 346 & 6.1   & 18.775 & 7.326  &6.2 \\
3 & 353 & 12.8  & 10.923 & 3.690  &5.2 \\
4 & 352 & 20.2  & 21.228 & 4.107  &10.2 \\
5 & 356 & 16.9  & 13.907 & 3.366  &8.7 \\
6 & 350 & 6.5   &  8.295 & 4.491  &6.7 \\
7 & 361 & 26.1  & 16.345 & 2.921  &8.2 \\
8 & 356 & 11.9  & 10.228 & 3.628  &7.7 \\
9 & 355 & 7.7   &  9.999 & 4.675  &6.4 \\
10& 351 & 3.9   &  4.053 & 4.234  &7.1 \\
11& 356 & 5.3   &  6.360 & 4.401  &5.0 \\
12& 346 & 2.2   &  4.562 & 6.591  &4.8 \\
13& 368 & 16.2  & 17.004 & 4.136  &8.2 \\
14& 364 & 10.9  & 16.622 & 5.194  &7.2 \\
15& 356 & 5.1   &  7.723 & 4.986  &6.8 \\
16& 364 & 7.4   & 18.976 & 5.988  &8.9 \\
17& 362 & 2.2   &  5.579 & 6.573  &4.0 \\
18& 382 & 12.9  & 18.487 & 4.825  &8.4 \\
19& 369 & 1.8   &  4.143 & 7.349  &3.6 \\
20& 370 & 2.6   &  4.882 & 5.491  &5.0 \\
21& 390 & 11.2  & 29.410 & 6.422  &9.3 \\
22& 392 & 2.8   &  8.089 & 6.738  &4.2 \\
23& 412 & 29.4  & 15.150 & 2.756  &7.8 \\
 \hline
\end{tabular}
\end{center}
\end{table}

\section[]{Observational properties of giant ellipticals}\label{obs}
We use the sample of 23, $z<0.3$, supermassive elliptical galaxies, 
selected by Bernardi et al. (2006, 2008) from the SDSS database 
on the basis of their large velocity dispersions:  $\sigma>$330 km~s$^{-1}$. 
Surface brightness fits and analyses of the SDSS and HST-based 
light profiles are presented in Hyde et al. (2008).  
An overview of the galaxy properties that we need for this study 
is gathered in Table \ref{tab:data}.

For all the objects in our sample, the observed surface brightness 
distribution was fit to de Vaucouleurs (DeV) and Sersic (S) profiles: 

\begin{equation}
 I^{DeV}(R) = I_0^{DeV}\, \exp\Bigl[-7.67\,[(R/r_{e}^{DeV})^{0.25}-1] \Bigr],
\end{equation}
 \begin{equation}
 I^S(R) = I_0 ^S\, \exp\Bigl[b_n\,[(R/r_{e} ^S)^{{1}\over{n}}-1] \Bigr],
\end{equation}
where $I_0$ and $r_{e}$ are the central surface brightness and the 
projected half-light radius ($r_e$ is given Table~\ref{tab:data}).  
(We use the parameters provided by Hyde et al. 2008, rather than 
those output from the SDSS database.  Hyde et al. also provide 
fits to Bulge/Disk decompositions, which we 
do not use here.)
We use $\gamma$ to denote the mass-to-light ratio, and we 
define the surface mass density $\Sigma(R) \equiv \gamma\, I(R)$, 
then the deprojected density, computed by inverting the Abel equation, 
is 
\begin{equation}
 \rho_{\star}(r) = -\frac{1}{\pi} \int_r^{\infty} \frac{d\Sigma(R)}{dR} 
                                               \frac{dR}{\sqrt{R^2-r^2}}.
\end{equation}
The mass within (a sphere of radius) $r$ is 
\begin{equation}
 M_*(<r) = 4\pi \int_0^r dx\,x^2\,\rho_{\star}(x) .
\end{equation}
Note that we could have defined the analogous quantities for the light, 
then $\rho_*(r) = \gamma \rho_L(r)$ and $M_*(<r) = \gamma\, L(<r)$.  
The quantity of most interest in this paper is the total stellar mass:
 $M_*\equiv M_*(\infty)\equiv \gamma \,L(\infty)$.  
We describe how we estimate it in the next section.  

The other observed quantity is the average velocity dispersion $\sigma_0$
within the SDSS fiber, which has a diameter (2$r_{ap}$)
of 3 arcsec, that corresponds to about 5-10 kpc according to the distance of our galaxies. 
(Note that here we do not use $\sigma$ aperture corrected to $r_e/8$
reported by Bernardi et al. 2006, 2008.) The value of $\sigma_0$ is related to the line-of-sight 
velocity dispersion of the object, weighted by the surface-brightness 
profile:
\begin{equation}
 \sigma_0^2 =  {{2\pi}\over{L(r_{ap})}} \int_0^{r_{ap}} 
                   \sigma_{los}^2(R)\, I(R)\, R\,dR
\label{eq:sigma2ap}
\end{equation}
where 
$L(r_{ap}) = 2\pi  \int_0^{r_{ap}} I(R)\, R\, dR$ and $\sigma_{los}^2(R)$ 
is the velocity dispersion at projected distance $R$ from the center.  

The quantity $\sigma_0$ is related to the gravitational potential, 
and hence to the total mass of the system, as follows.  

\section{Jeans' equation analysis}\label{jeans}
In what follows, we provide an estimate of the total mass, and the 
fraction of this which is in stars, for the objects in our sample.  
Our analysis assumes that the objects -- both the stellar and the 
dark matter components -- are spherical, with constant stellar 
mass-to-light ratios, and no anisotropic velocities.  This is 
extremely idealized:  Bernardi et al. (2008) have argued that 
many of these objects are likely to be prolate objects viewed along 
the long axis.  We discuss this more complex case in an Appendix, but 
since none of our conclusions are sensitive to this, we have kept 
the simpler (spherical, isotropic) model in the main body of the 
text.   Bernardi et al. also argue that some of these objects 
may be rotating -- an effect we do not include in our analysis.  
We have not removed such objects from our analysis, 
since it is interesting that they appear to show 
similar scalings as the objects which are not rotating.  

\subsection{Spherical symmetry and isotropic velocities}
The 1-D Jeans equation in spherical symmetry (Binney \& Tremaine 1987) 
relates the radial velocity dispersion $\sigma_r(r)$ to the mass 
distribution: 
\begin{equation}
 \frac{d[\rho_\star(r) \sigma_r^2(r)]}{dr} 
  + 2\beta(r)\, \frac{\rho_\star(r) \sigma_r^2(r)}{r} = 
  -\rho_\star(r)\, \frac{GM(<r)}{r^2},
\label{eq:jeans}
\end{equation}
where $\rho_\star$ is the density of the stellar component at $r$, 
$M(<r)$ is the total mass ($M_\star+M_{DM}$) within $r$ and $\beta$ is 
the anisotropy profile.  Studies of nearby elliptical galaxies indicate 
that the tangential anisotropy within the half-light radius is negligible 
(e.g. Matthias \& Gerhard 1999;  Gerhard et al. 2001; Koopmans \& Treu 2003; 
Koopmans et al. 2009).  Therefore we set $\beta=0$ in what follows -- 
the Appendix shows how our results are modified if $\beta\ne 0$.  
Thus, equation~(\ref{eq:jeans}) implies 
\begin{equation}
 \sigma_r^2(r) = \frac{G}{\rho_\star(r)} 
                  \int_r^\infty \frac{\rho_\star(r)M(<r)}{r^2}\,dr.
\label{eq:sigma2}
\end{equation}
An observer only measures the projection along the line of sight, 
$\sigma_{los}(r)$, which is given by 
\begin{equation}
\sigma_{los}^2(r) =  \frac{2}{\Sigma(r)} \int_r^\infty 
   \frac{\rho_\star(R) \sigma_r^2(R)}{\sqrt{R^2-r^2}}\, R\, dR.
\label{eq:sigma2los}
\end{equation}
Inserting equation~(\ref{eq:sigma2}) in~(\ref{eq:sigma2los}), 
and this in~(\ref{eq:sigma2ap}) shows how the total dynamical mass 
is related to the observed light profile and $\sigma_0$.

\begin{figure*}
 \centering
\includegraphics[width=168mm]{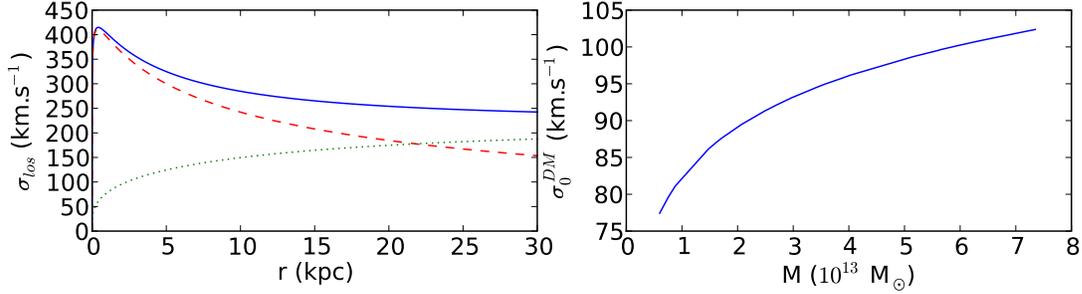}  
 \caption{\textit{Left}: Radial velocity dispersion velocity profile 
for a massive elliptical galaxy having stellar mass $M_*=10^{12}M_\odot$ 
surrounded by a dark matter halo of mass $M_{vir}=30M_*$.  The radial 
distributions of these two components are described in the main text.  
The dotted line (green) represents the dark matter component, 
the dashed line (red) the stellar component (for a de Vaucouleurs profile), and the solid line (blue) 
the total profile. For such objects, the dark matter does not 
contribute significantly within the first kpc.  
\textit{Right}: Central velocity dispersion measured within a fiber 
  of projected radius 7~kpc ($\sigma_0^{DM}$), for the dark matter 
  component only, as $M_{vir}$ is increased. }
\label{fig:disp}
\end{figure*}

\subsection{Insignificance of dark matter in $\sigma_0$}

In what follows, we estimate the total mass within some radius $r$ 
as the sum of the stellar mass (obtained from the observed light 
profile with the assumption of constant mass-to-light ratio) and 
that of the dark matter, for which we use the fitting formula 
of Navarro et al. (1996; hereafter NFW).

For an NFW profile with total mass $M_{vir}$, the virial radius is 
\begin{equation}
 \frac{r_{vir}}{\rm kpc} = 548\,\left(\frac{M_{vir}}{10^{13} M_\odot}\right)^{0.33}.
\end{equation}
The mass within some $r<r_{vir}$ is given by 
\begin{equation}
 M_{DM}(<x) = M_{vir}\,\frac{\log(1+x)-\frac{x}{1+x}}{\log(1+c)-\frac{c}{1+c}},
\label{eq:nfw1}
\end{equation}
where $x = r/r_s = cr/r_{vir}$:  
$r_s$ is a characteristic scale length for the halo, and the final 
equality defines the concentration parameter 
\begin{equation}
 c \equiv \frac{r_{vir}}{r_s} 
   \approx 9.3\, \left(\frac{M_{vir}}{10^{13} M_\odot}\right)^{-0.13}.  
\label{eq:nfw2}
\end{equation}
Note that massive halos are less concentrated.  
For small $x$, 
\begin{equation}
 \frac{M_{DM}(<x)}{M_{vir}} \approx \frac{x^2/2}{\log(1+c)-\frac{c}{1+c}};
\label{eq:nfw1}
\end{equation}
for $c=9.3$, this is $x^2/2.86$.  

We model the total mass of each galaxy in our sample as a superposition 
of the deprojected de Vaucouleurs or Sersic profile having total mass $M_*$ and an 
NFW halo with a total mass $M_{vir}$.  
This represents a compromise:  adiabatic contraction arguments suggest 
that the dark matter should become more centrally concentrated than a 
pure NFW profile, as the gas which formed the stellar component shrinks 
towards the center -- we are ignoring this effect (see 
Padmanabhan et al. 2004; Schulz et al. 2010; Treu et al. 2010 for 
recent analyses of other samples in which this effect is included).  
On the other hand, observations suggest that galaxies have a cored 
rather than a cuspy halo (e.g. Salucci et al. 2007) -- even an 
uncontracted NFW profile is too steep.

\begin{table}
\caption{Results of the modelling. Contribution of the dark matter component on the total dispersion velocity (column 2).  Dynamical mass computed by solving the Jeans equation (column 3-4). Photometric stellar mass using the model of Maraston (column 5-6).}
\begin{center}
\begin{tabular}{rrrrrrr}
  \hline
\# &${{\sigma_0^{tot}-\sigma_0^{\star}}\over{\sigma_0^{tot}}}$&$M^{dyn,DeV}_{\star} $&$M^{dyn,S}_{\star} $ & $M^{SP,DeV}_{\star} $& $M^{SP,S}_{\star} $\\
&($\%$)&($10^{11}$M$_{\odot}$) &   ($10^{11}$M$_{\odot}$)&($10^{11}$M$_{\odot}$) &   ($10^{11}$M$_{\odot}$)   \\
 \hline

1	&   5 &    9.5 & 11.2  & 5.5 &  8.5 \\
2	&   3 &    7.6 & 8.3   & 7.4 & 13.2 \\
3	&   5 &  15.0  & 14.4  &14.5 & 13.2 \\
4       &   8 &  21.7  & 22.1  &11.7 & 12.3 \\
5	&   6 &  19.7  & 19.6  &14.5 & 12.6 \\
6	&   2 &    8.3 & 9.0   &23.4 & 26.9 \\
7	& 10 &  28.7   & 25.6  &24.0 & 18.2 \\
8	&   5 &  14.5  & 14.1  & 6.6 &  6.0 \\
9	&   4 &    9.7 & 10.3  & 5.2 &  6.0 \\
10	&   1 &    5.2 & 5.2   & 7.9 &  7.9 \\
11	&   2 &    6.9 & 7.3   & 2.2 &  2.4 \\
12	&   1 &    3.0 & 3.1   & 2.6 &  3.6 \\
13	&   6 &  20.4  & 20.7  &10.0 & 10.2 \\
14	&   4 &  14.2  & 15.0  &18.2 & 22.9 \\
15	&   3 &    6.8 & 7.5   & 3.4 &  4.3 \\
16	&   3 &  10.1  & 12.4  & 6.6 & 11.0 \\
17	&   2 &    3.1 & 3.6   & 0.9 &  1.4 \\
18	&   5 &  18.1  & 19.6  & 8.1 & 10.0 \\
19	&   1 &    2.7 & 2.7   & 2.9 &  3.6 \\
20	&   1 &    4.0 & 4.6   & 6.9 &  9.5 \\
21	&   3 &  17.3  & 19.8  &25.7 & 42.7 \\
22	&   1 &    4.7 & 5.7   & 7.1 & 12.3 \\ 
\hline

\end{tabular}
\end{center}
\label{tab:results}
\end{table}%

To proceed further, we assume that :
\begin{equation}
M_{vir} = 30M_{\star},
 \label{eq:mvir30}
\end{equation}
following Shankar et al. (2006).  This makes 
\begin{equation}
 M(<r) %= M_*(<r) + M_{DM}(<r) 
       = M_* \left(\frac{M_*(<r)}{M_*} + 30\,\frac{M_{DM}(<r)}{M_{vir}}\right).
 \label{eq:shankar30}
\end{equation}
Notice that big galaxies are likely to have some scatter around this number
(i.e. 30). In fact for smaller 
values, our conclusion that the stars dominate the mass
will be stronger.
We demonstrate that the mass within
$r_{ap}$ is dominated by the stars, not the dark matter.
For $M_{vir}\approx 10^{13}M_\odot$, $r_s = r_{vir}/c \approx 60$~kpc, and the NFW mass within $r_{ap}$ is approximately 
$30M_*\,(r_{ap}/r_s)^2/2.86$, whereas the stellar mass within $r_{ap}$ is 
slightly less than $M_*/2$.  Since $r_{ap}/r_s\ll 1$, the total mass 
within $r_{ap}$ is dominated by the stellar mass.  Consequently,  $\sigma_{0}$ is also 
dominated by the stellar mass. 

 We show this explicitly in 
Figure~\ref{fig:disp}. The left hand panel shows $\sigma_{los}(r)$ , obtained from equation \ref{eq:sigma2}, \ref{eq:sigma2los},
for an elliptical galaxy with $M_\star=10^{12}M_\odot$ and $r_e= 6$~kpc.
We see that the contribution from the dark matter component is always 
much smaller than that of the stellar component ($\sigma_{los}^{DM}\ll \sigma_{los}^\star$).

Then we show that $\sigma_0^{DM}$ 
is always negligible compared to $\sigma_0$ observed (right hand panel).
Since $\sigma_{0}^2 = \sigma_0^{\star \ 2} + \sigma_0^{DM\ 2}$,  within our assumptions, the measured value of $\sigma_0$ 
provides a good estimate of $M_*$, without being contaminated by the dark 
matter component. 
The effect of DM is small, we can evaluate it and correct the estimation of $\sigma_0^{\star}$ by assuming equation \ref{eq:mvir30}.
On average the dark matter component contributes to less than $5\%$ to the total velocity dispersion (Table \ref{tab:results}).

\begin{figure}
 \centering
 \includegraphics[width=84mm]{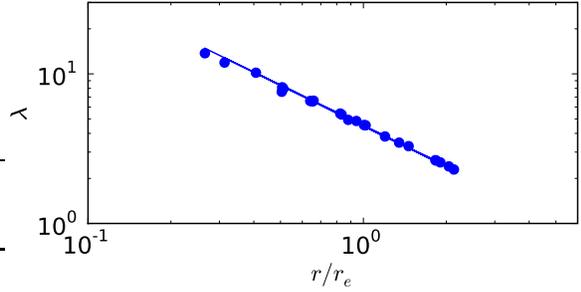}  	 
 \caption{Constant of proportionality $\lambda$ between $M_*$ and 
  $r_{ap}\sigma_{0}^2$ (equation~\ref{eq:lam}), 
 where $M_*$ is computed by solving the Jeans equation (see text).}
 \label{fig:lambda}
\end{figure}

\begin{figure}	
\centering	
\includegraphics [width=74mm]{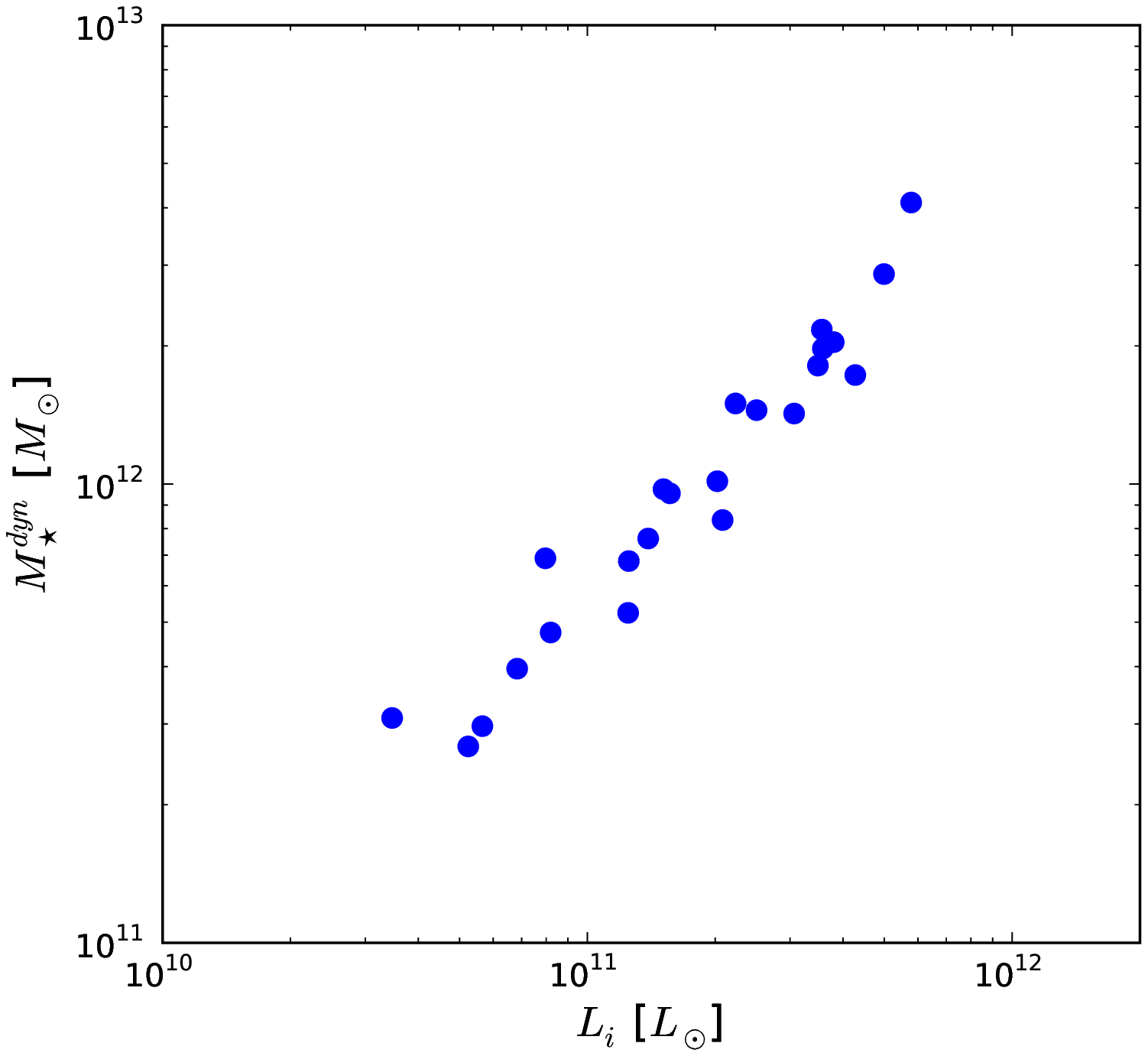}
\includegraphics [width=74mm]{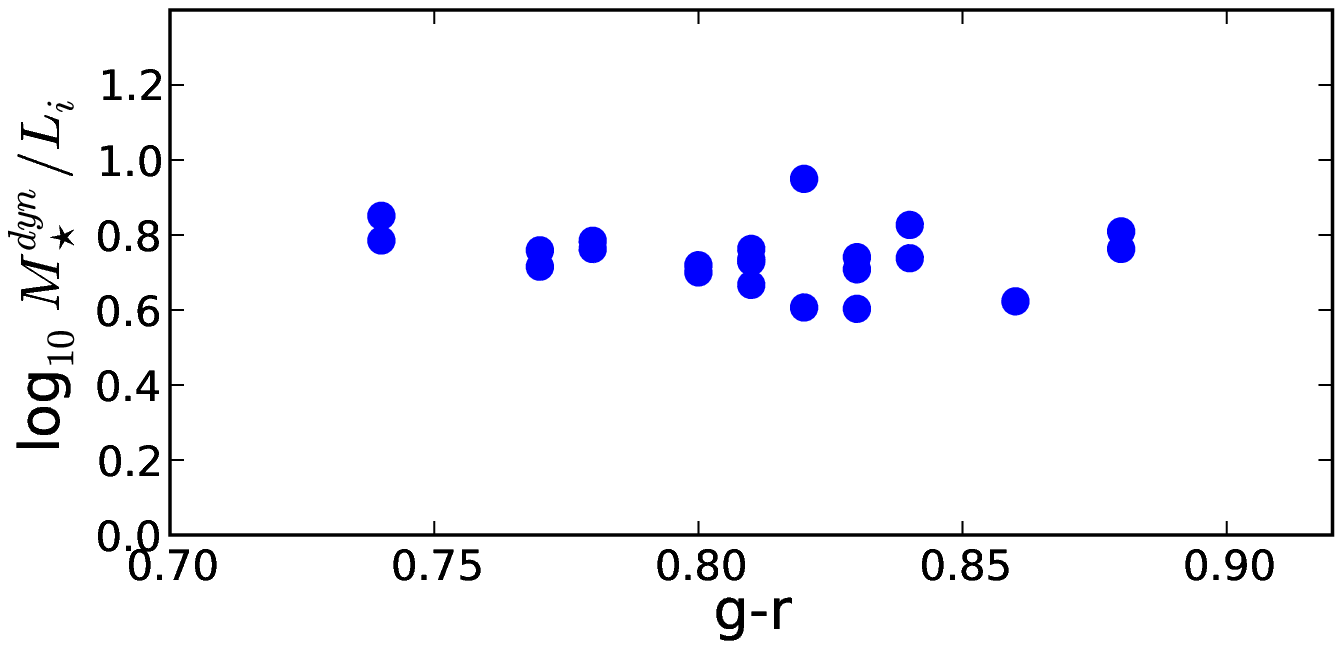}
\caption{\textit{Top}: Dynamical mass versus luminosity in the i-band. \textit{Bottom}: stellar mass-to-light ratio (computed from the dynamical mass) versus color g-r.}	
\label{fig:Li}	
\end{figure}

Now we build a new stellar mass estimator that exploits 
the quantity $\sigma_{0}$.  The Jeans equation implies that 
the quantity $r_{ap} \sigma_{0}^2$ should be proportional to a function of $r_{ap}/r_e$.  
If we set 
\begin{equation}
 M_{\star} = \lambda \, (r_{ap} \sigma_{0}^2),
 \label{eq:mdyn}
\end{equation}
the actual value of $\lambda$ plotted in Figure~\ref{fig:lambda} can be fitted by:
\begin{equation}
 \lambda = 4.5\,\left(\frac{r_{ap}}{r_e}\right)^{-0.9}.
 \label{eq:lam}
\end{equation}
As a check, note that if the power-law above had slope $-1$, then 
$M_*\propto r_e\sigma_{0}^2$;  this is the scaling that is usually 
assumed.  The zero-point of the $M_*\propto r_e\sigma_{0}^2$ relation, 
4.5, is close to that commonly assumed for a pure de Vaucouleurs profile
(e.g. Bernardi et al. 2010a).  
It is also close to that derived from analyses which assume adiabatic 
contraction (Padmanabhan et al. 2004).  

The top panel of Figure \ref{fig:Li} shows the dynamical stellar mass computed as described above, versus the luminosity in the i-band.
The bottom panel shows the corresponding mass-to-light ratio versus color (g-r).

\section{Comparison with stellar mass estimates from photometry}\label{ssp}

We estimate the stellar masses $M_*^{\rm SP}$, from the stellar population (hereafter SP) model of Maraston et al. (2009) (available at www.maraston.eu).  This is a two-component model made by a major metal-rich population and a low percentage ($3\%$) of an old and metal-poor population with the same age. Moreover, an improvement in the spectra of K-giants is included through the empirical stellar spectra of Pickles (1998). The initial mass function in the Maraston et al. (2009) paper was a Salpeter (1955) and we did not modify this assumption here. We use these models because they trace the color evolution of Luminous Red Galaxies (LRGs) in SDSS from redshift 0.1 to 0.6 much better than previous attempts did (see Maraston et al. 2009). 

We obtain the stellar masses through Spectral Energy Distribution (SED) fitting, as in Maraston et al. (2006), e.g., by using the Maraston et al. 2009 templates in the Hyper-Z code (Bolzonella et al. 2000)\footnote{We checked that had we fitted a composite model with identical characteristics, but obtained with the Maraston (2005) SP models, i.e. the standard models based on the Kurucz (1979) model atmospheres, we would have obtained the same results.}. We computed the stellar masses using both de Vaucouleurs and Sersic  magnitudes. The first set were obtained by fitting the ugriz de Vaucouleurs magnitudes
available from the SDSS database, while the second set was computed
rescaling the SDSS de Vaucouleurs magnitudes by the difference between
the i-band Sersic and de Vaucouleurs magnitude observed by Hyde et al. (2008).
This is a good approximation since color gradients are small. 

\begin{figure*}
 \centering
\includegraphics[width=168mm]{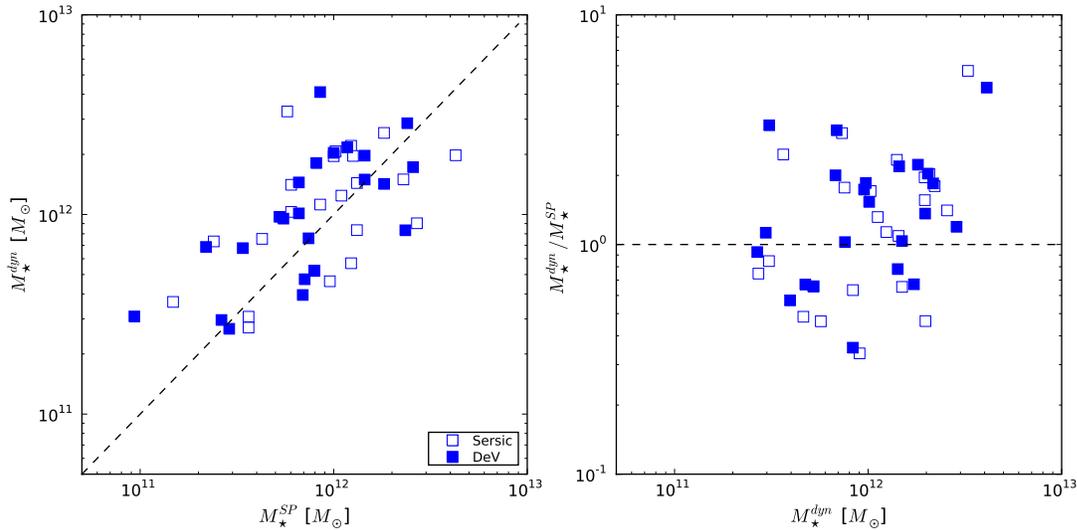}
 \caption{\textit{Left}:  Dynamical masses from the Jeans equation versus stellar 
  masses from the composite stellar population models by Maraston et al. (2009). 
  Dashed line shows $M_{\star}^{\rm dyn}=M_{\star}^{\rm SP}$.  
   \textit{Right}: Ratio of $M_{\star}^{\rm dyn}/M_{\star}^{\rm SP}$ as a function of 
  $M_{\rm dyn}$. }
\label{fig:MdynMvis}
\end{figure*}

The panel on the left of Figure~\ref{fig:MdynMvis} shows our 
$M_*^{\rm dyn}$ estimates versus those from the stellar population 
model.  Note that $M_\star^{\rm dyn}=M_\star^{\rm SP}$, 
the stellar masses as obtained with these composite models are in agreement with the dynamical masses. 

\begin{figure*}
 \centering
 \includegraphics[width=168mm]{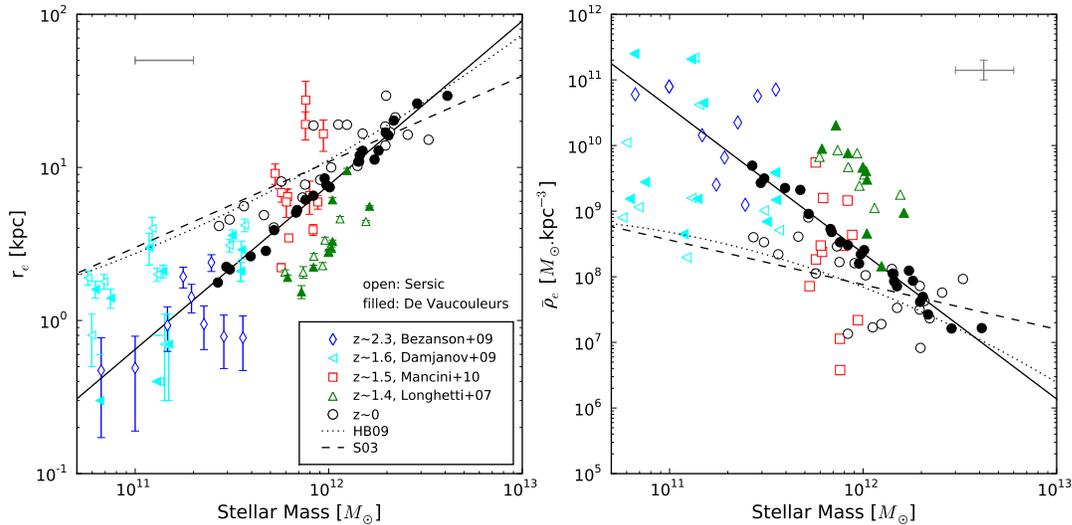}
 \caption{\textit{Left}: Correlation between the characteristic radius, 
  $r_e$, and the stellar mass $M_*$. The black circles correspond to our sample (where the mass is estimated from the Jeans equation 
  analysis). Open and filled symbols show the objects modelled using a Sersic or de Vaucouleurs profile, respectively.  
  The dotted line shows the relation for the full sample of early-types 
  from Hyde \& Bernardi (2009) and the dashed line is from 
  Shen et al (2003).  
  \textit {Right}: Correlation between the mean stellar density within $r_e$ and $M_*$. }
\label{fig:comp1}
\end{figure*}

\section{Structural properties of giant ellipticals}\label{reM}
We now analyze the correlation between size and mass in our objects. 
The left hand panel of Figure~\ref {fig:comp1} shows the correlation 
between $r_e$ and $M_*$ using our dynamical estimates of $M_*$.  
The solid line shows the direct fit (i.e. $\langle r_e|M_*\rangle$)
\begin{equation}
 \frac{r_e}{\rm kpc} = 7.2 \,
              \left(\frac{M_{\star}}{10^{12} M_\odot}\right)^{1.07\pm 0.04}.
\end{equation}

This relation is significantly steeper than the 
$\langle r_e|M_*\rangle \propto M_*^{0.6}$ that is usually reported 
(e.g. Hyde \& Bernardi 2009).  However, if we recall that these 
objects all have large $\sigma$, then the relevant comparison is to 
the relation at fixed $M_*$ {\em and} $\sigma$.  This slope is close 
to 0.9 (Bernardi et al. 2008).  In this case, it is plausible that 
our (now only slightly) steeper slope is due to the fact that our 
$M_*$ estimator is less noisy. 

In fact, the small scatter in our relation can also be understood 
in these terms.  Equations~(\ref{eq:mdyn}) and~(\ref{eq:lam}) show 
that our dynamical estimate of $M_*$ is proportional to 
$(r_e/r_{ap})^{0.9}\,r_{ap}\sigma^2 = (r_{ap}/r_e)^{0.1}\, r_e\sigma^2$.  
Figure~\ref{fig:lambda} shows that there is little scatter around 
this relation.  However, our sample has essentially fixed $\sigma$, 
and $r_{ap}/r_e$ has only a small scatter, so Figure~\ref{fig:comp1} 
is almost a plot of $r_e$ vs $r_e$, with the zero-point being 
set by the mean $\sigma$ and $r_{ap}/r_e$ values in the sample.  
This is why the slope is close to unity. Of course, if the stellar distribution 
were not homologous, or the dark matter was a significant fraction 
of the total mass, then $r_{ap}/r_e$ need not have small scatter.

\begin{figure*}
 \centering

\includegraphics[width=168mm]{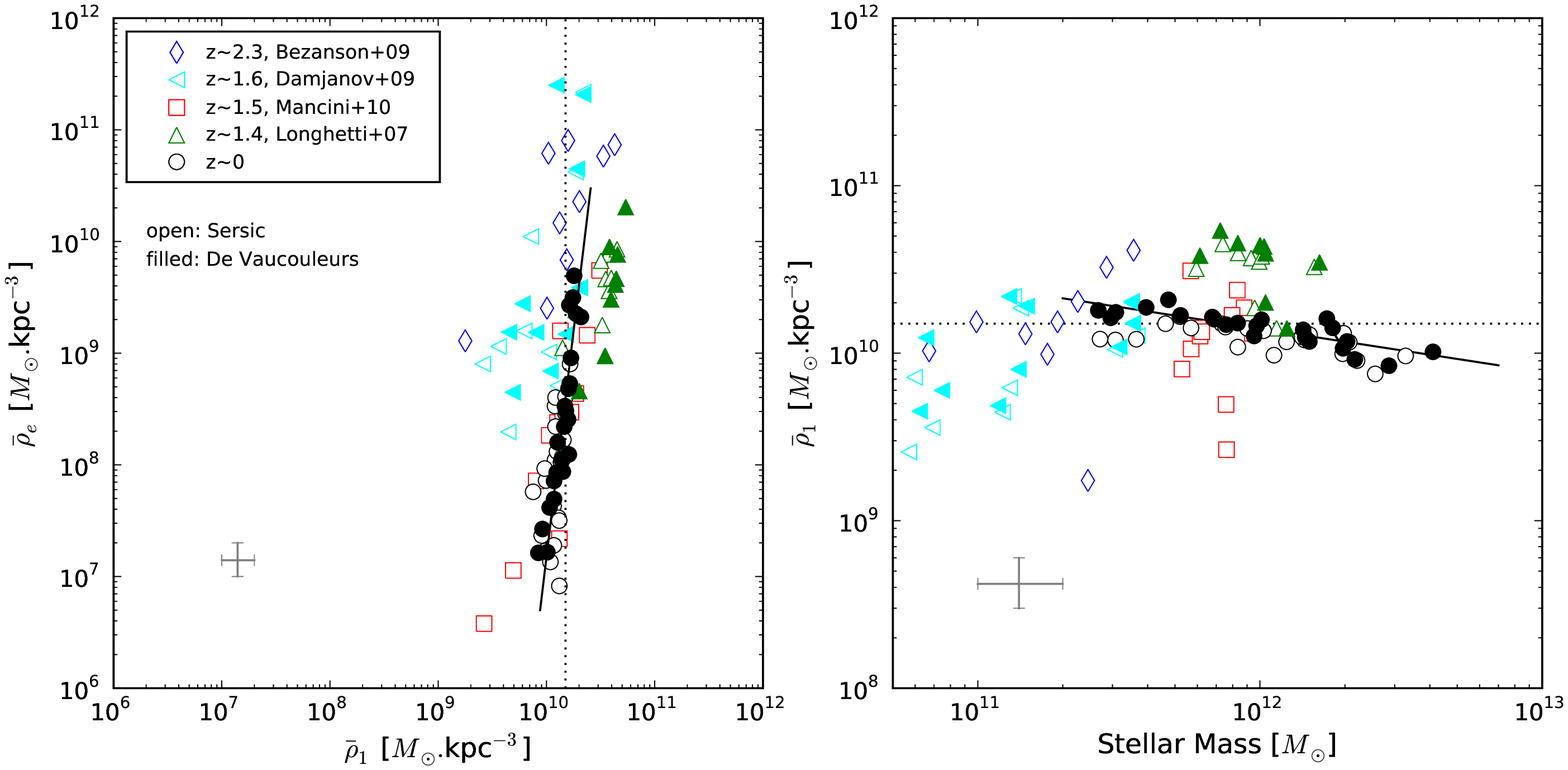}
 \caption{\textit{Left}: Stellar mass density within r$_e$ versus
  the stellar mass density within 1 kpc. \textit{Right}: Correlation 
  between stellar 
  mass density within 1 kpc and the total stellar mass.  In both panels, 
  open symbols are from a Sersic profile, while filled symbols are from a 
  de Vaucouleurs profile. }
\label{fig:comp2}
\end{figure*}

The diamonds, squares, triangles show these same relations, but now 
obtained from a sample of $z\sim 2.3$ objects by Bezanson et al. (2009), 
Damjanov et al. (2009), Mancini et al. (2010) and Longhetti et al. (2007), 
respectively. A number of studies have noted that the 
$z\sim 2$ objects are significantly smaller than $z\sim 0$ objects of the 
same $M_*$ (dotted or dashed curves).  Note that
Bezanson et al., Mancini et al., Longhetti et al. and Damjanov et al. computed their stellar masses with different
IMF (Kroupa, Chabrier, Salpeter and Baldry \& Glazebrook) and different stellar population models
(Bruzual \& Charlot 2003 and Maraston 2005). The fit from 
Hyde \& Bernardi (2009) and Shen et al. (2003) were computed using a 
Chabrier IMF. 
In order to compare these different samples we recalibrated the masses 
to a Salpeter IMF (as used in Maraston et al. 2009) using these scale-factors:
$M_\star^{Salpeter} = 1.6 M_\star^{Kroupa} = 1.78 M_\star^{Chabrier}= 2 M_\star^{B\&G}$.
In addition, to account for differences in stellar population models
we rescaled the Bezanson et al. data multiplying their stellar masses
by 0.7 (as e.g. in Mancini et al. 2010, see also Muzzin et al. 2009)
and by 0.5 for the Damjanov et al. stellar masses.

The evidence for small sizes at $z\sim 2$ is not uncontested.  
Mancini et al. (2010) have argued that neglect of low surface brightness 
features will bias $r_e$ to small values (while the bias in $M_*$ is small --
C. Mancini private communication).  Accounting for this effect in a sample at $z\sim 1.5$ yields the open squares 
in Figure~\ref{fig:comp1}.  Evidently, these objects are not small 
for their $M_*$ compared to $z=0$ objects.  If both Mancini et al. 
and Bezanson et al. are correct, and both probe the massive end of 
the population at their respective mean redshifts, then there must 
have been significant evolution in size and stellar mass between $z=2.3$ 
and $z=1.5$.  

In view of this discussion, it is remarkable that the $z=2.3$ 
compact galaxies appear to trace the small size and $M_*$ end of 
the relation we find at $z\sim 0$, for fixed $\sigma$ (solid line).  
The $z\sim 1.5$ sample of Mancini et al. (2010) is confined to a 
narrower range of $M_*$, making it difficult to define a relation.  
However, at this $M_*$, the difference between the solid line 
$(\sigma\sim 400$~km~s$^{-1}$) and the others (bulk of early-type 
population) is a factor of two or less.  It will be interesting if 
future measurements show the $z=2.3$ objects to have 
$\sigma \sim 400$~km~s$^{-1}$, and even more so if this is also true 
for the most compact of the objects in the Mancini et al. sample.  

We now turn to a slightly different version of the correlation between 
size and mass, namely that between average density and mass. For this 
purpose, it is useful to define 
\begin{equation}
 \bar{\rho}_{r} = \frac{3}{r^3} \,\int_0^r \rho_\star(r)\, r^2\,dr,
\end{equation} 
the average density within $r$. 
In what follows, we will pay special attention to $\bar{\rho}_e$ 
and $\bar{\rho}_1$:  
the mean density within the de Vaucouleurs radius $r_e$, 
and within 1 kpc, respectively.  
Whereas $\bar\rho_e$ can be thought of as a characteristic density, 
$\bar{\rho}_1$ is more like the central density (recall that, for this 
sample, $r_e\approx 10$~kpc).  

The right-hand panel of Figure~\ref {fig:comp2} shows the 
correlation between $\bar{\rho}_e$ and $M_*$, using the same 
format as for the panel on the left.  Fitting to the relation 
defined by our dynamically estimated $M_*$ yields 
\begin{equation}
 \frac{\bar{\rho}_e}{M_\odot\,{\rm kpc}^{-3}} = 2.3 \times 10^{8}  \,
   \left(\frac{10^{12} M_\odot}{M_{\star}}\right)^{2.22\pm 0.04} .
\end{equation}

This characteristic density is a sharply declining function of 
stellar mass -- the decline is significantly steeper than previously 
reported for a sample which includes the full range of early-types 
(e.g. Bernardi et al. 2003; Hyde \& Bernardi 2009).  
Following our discussion of the 
$r_e-M_*$ relation above, the more relevant comparison may be with 
the $\bar\rho_e-M_*$ relation at fixed $\sigma$, for which the 
slope is $-1.8$ (Bernardi et al. 2008).  
Our current estimate is slightly steeper, perhaps 
because our mass estimate is less noisy.  
To see this, note that we could have derived the slope from the fact 
that 
 $\bar\rho_e\propto M_*/r_e^3 \propto M_*^{-2.21}$,  
where the final expression uses the fact that the scatter between 
$r_e$ and $M_*$ is small (which we argued was a consequence of 
the fact that our sample has only a small range of $\sigma$, 
a small range in $r_e/r_{ap}$, and that Figure~\ref{fig:lambda} has 
small scatter).  
We note that, while the $z=2.3$ objects have $\bar\rho_e$ orders 
of magnitude larger than the bulk of the $z=0$ objects of the same 
$M_*$ (compare diamonds with dashed or dotted curves), 
they are only slightly denser than $z=0$ objects of the same $M_*$, 
if such objects had $\sigma\sim 400$~km~s$^{-1}$ (extrapolate 
solid line to small $M_*$).  

The steepness of this relation stands in stark contrast to the 
relation between $\bar\rho_1$ and $M_*$.  The right-hand panel of 
Figure~\ref{fig:comp2} shows that this relation is very shallow. We find 
\begin{equation}
 \frac{\bar{\rho}_1}{M_\odot \,{\rm kpc}^{-3}} = 1.8 \times 10^{10} 
  \left(\frac{10^{12} M_\odot}{M_{\star}}\right)^{0.25\pm0.05}.  
\end{equation}
While the mass varies by 2 orders of magnitude (from $10^{11}$ $M_{\odot}$ 
to $5 \times 10^{12}$ $M_{\odot}$), the central density remains constant 
at about $1.8 \times 10^{10}$  $M_{\odot}\,{\rm kpc}^{-3}$. 

The left-hand panel of Figure~\ref{fig:comp2} shows another way 
of presenting this information:  while $\rho_e$ varies by 
3 orders of magnitude, $\bar\rho_1$ varies by less than a factor of 2.  
We find 
\begin{equation}
 \frac{\bar{\rho}_1}{M_\odot \,{\rm kpc}^{-3}} = 1.3 \times10^{9} \,
  \left(\frac{\bar{\rho}_e}{M_\odot \,{\rm kpc}^{-3}}\right)^{0.12\pm0.04}.
\end{equation}
Thus the central density is approximately constant for all the 
galaxies of our sample.

It is quite remarkable that they show the same relation as in the samples of Mancini et al. and Longhetti et al. at $z \sim 1.5$  as well as of Bezanson et al. (2009) at $z\sim 2.3$.  Although $\bar\rho_e$ 
is orders of magnitude larger in the $z\sim 2.3$ sample than at $z=0$, 
$\bar\rho_1$ is different by only a factor of two or so.

\section{Discussion and conclusions}
We analyzed a sample of 23 giant elliptical galaxies. They are 
very massive ($M>10^{11} M_{\odot}$) with large central velocity dispersions  
$\gsim 330$~km~s$^{-1}$, which suggests old stellar populations (Bernardi et al. 2005).  
We estimated dynamical masses for each of these systems by using the 
observed light profiles and central velocity dispersions, 
by solving the Jeans Equation, under the assumptions of spherical 
symmetry, no tangential velocity dispersions, no radial dependence 
of the stellar mass-to-light ratio, and that each system has 30 times 
more dark matter than stellar matter within the virial radius, 
with the dark matter following an NFW profile (i.e. no accounting 
for adiabatic contraction).  

We found that the total stellar masses of these systems vary from 
about $10^{11} M_{\odot}$ to $5\times 10^{12} M_{\odot}$ 
(Table~\ref{tab:results}).   
In such models, the contribution of the dark matter modifies the central velocity dispersion by less 
than $5\%$ (Figure~\ref{fig:disp}).  Thus, for these objects, the 
observed $\sigma_0$ provides a good estimator of the luminous mass (equation~\ref{eq:lam}). 
We compared the masses we derive to estimates of the stellar mass from 
stellar population models (Figure~\ref{fig:MdynMvis}). We find good agreement using a composite model with high age and a small (by mass) metal-poor sub-component.  This model fits well the colors of Luminous Red Galaxies in SDSS.
A major result of this study is that we compute the mass-to-light ratio for massive elliptical galaxies. This ratio is roughly constant for all the sample (Figure \ref{fig:Li}), we find $M_\star/L\sim5 \pm 1$ in the i-band,  for  $0.7< g-r < 0.9$.

\begin{figure*}
 \centering
\includegraphics[width=84mm]{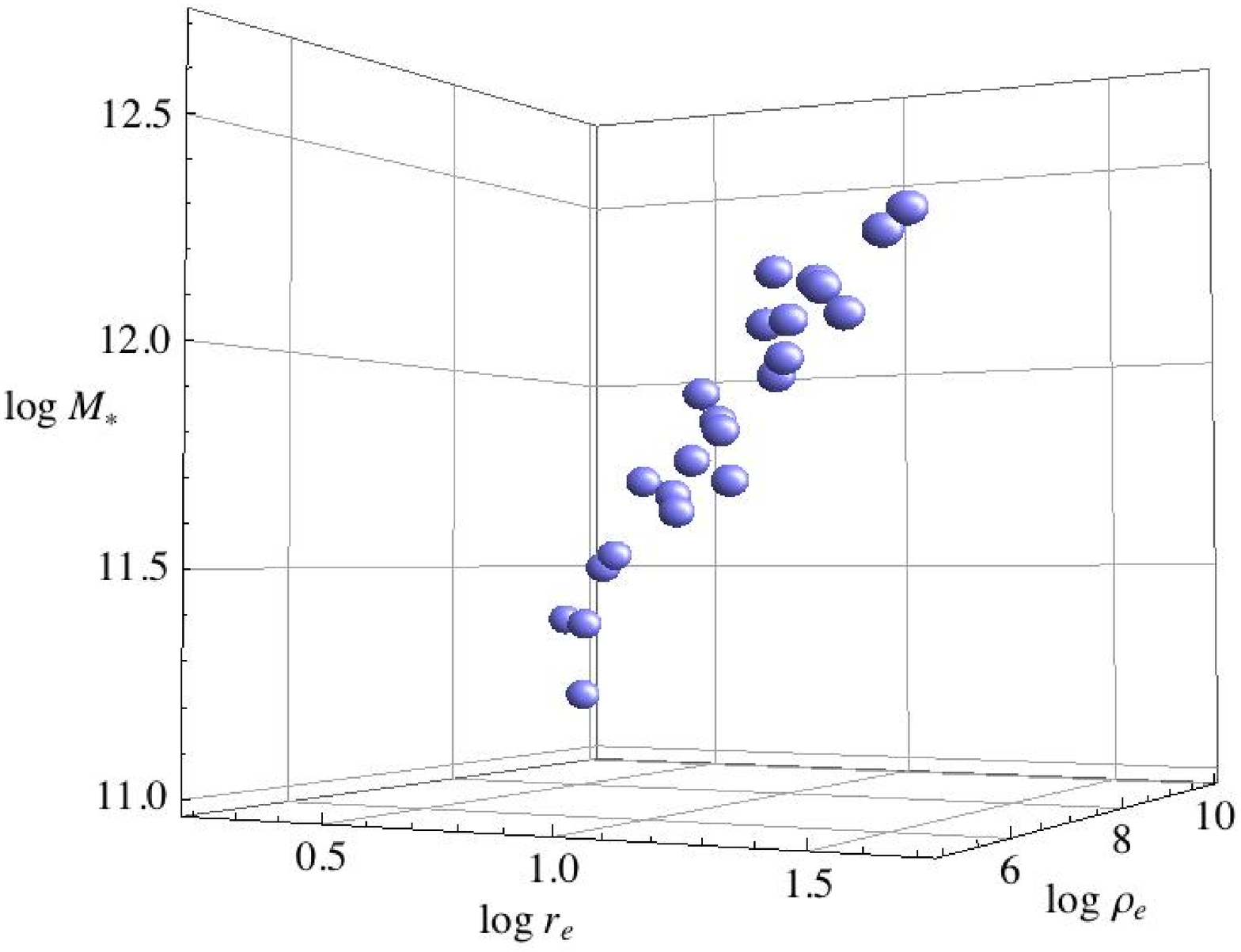}
\includegraphics[width=84mm]{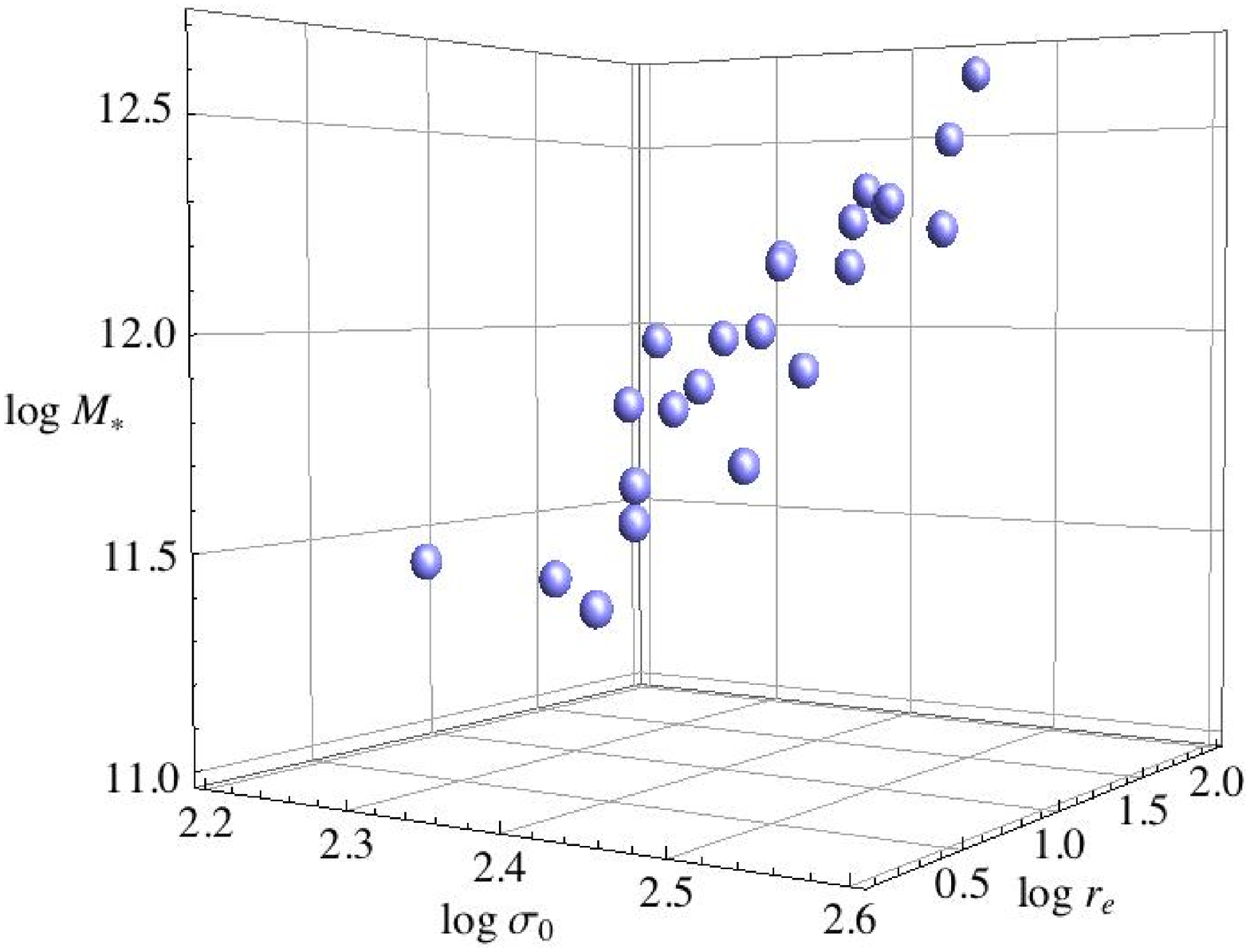}
 \caption{\textit{Left}: Relation between the stellar mass $M_\star$, 
          the characteristic density $\rho_e$, and the de Vaucouleurs 
          radius $r_e$. 
          \textit{Right}: Relation between $M_\star$, $\sigma_0$ and $r_e$.}
\label{fig:plot3d}
\end{figure*}

We also studied different ways of presenting the correlation between 
size and mass (Figure~\ref{fig:comp1}).  The correlation we find, 
$r_e\propto M_*^{1.07}$, is slightly steeper than that reported 
by Bernardi et al. (2008).  We suggest that this is because our 
estimate of $M_*$ is less noisy.
For similar reasons, we find that the density within $r_e$ is a 
more strongly decreasing function of $M_*$ than reported by 
Bernardi et al. (2008).  We find 
  $\bar\rho_e\propto M_*/r_e^3 \propto M_*^{-2.2}$ 
compared to their $-1.8$. 

A notable result is that galaxies at z $\sim$ 2.3 (Bezanson et al.) and z $\sim$ 1.5
(Mancini et al.) appear to follow the same relation that we find at z $\sim$ 0.
If each sample correctly gather the most massive galaxies for each range of redshift,
the evolution between the size and the stellar mass should  be meaningful.

The small scatter associated with our dynamical estimator of $M_*$ 
means that, in the space of $r_e$, $\bar{\rho}_e$ and $M_\star$, 
the objects in our sample trace out a one-dimensional curve (although we have argued, as did 
Bernardi et al. 2008, that the $r_e\propto M_*$ scaling we find 
is consistent with the simplest virial theorem scaling, once we 
account for the fact that these objects have essentially fixed $\sigma$).  
This is also true in the space of $r_e$, $\sigma$ and $M_\star$. 
We show this explicitly in Figure~\ref{fig:plot3d}.  Had we 
used the stellar population-based estimate of $M_*$, these curves 
would have been broadened into a plane.  Since the ($r_e, \sigma_0, M_\star$)
projection is similar to that of the Fundamental Plane, our results 
suggest that scatter in the relation between $L$ and $M_*$, or 
uncertainties in estimating $M_*$ serve to enhance the impression 
of a plane rather than a curve.  

On the other hand, some of the decreased scatter in our analysis is 
due to our neglect of anisotropic velocity dispersion profiles.  
We explore this in the Appendix.  
Nevertheless, it is likely that the scaling relation 
between $r_e$ and $M_*$ of giant ellipticals is significantly steeper 
than for spirals (which have $r_e\propto M_{\star}^{1/2}$).  
Understanding why is a challenge for models in which ellipticals form 
from mergers of spirals.

Although $\bar\rho_e$ decreases strongly with $M_*$, the average density 
on smaller scales (we chose 1~kpc) is almost independent of $M_*$ 
(Figure~\ref{fig:comp2}).  Moreover, it is remarkably similar to 
that found by Bezanson et al. (2009), in their analysis of $z\sim 2.3$ 
galaxies.  
The mean density within 1 kpc seems to be independent of the redshift 
and of the mass on an object by object basis.  I.e., since $z\sim 2$, 
as these galaxies grew in mass, the mass in the inner kpc remained 
unchanged.  Understanding why is an interesting challenge.

Although this is most easily accomplished in models where the mass 
is added to the outer regions only (e.g. Lapi \& Cavaliere 2009, Cook et al 2009) 
-- so it is tempting to conclude 
that minor mergers were the dominant growth mode since $z\sim 2$ 
(e.g. Bezanson et al. 2009) -- there is a direct counterexample to 
this conclusion in the literature.  In numerical simulations of 
hierarchical structure formation, Gao et al. (2004) find that 
although the mass in the central regions of what becomes a massive 
cluster at $z=0$ has remained constant since $z\sim 6$, the particles 
which make up this mass changed dramatically as the objects assembled. 
This assembly occurred through a sequence of major mergers at $z>1$; 
with minor mergers beginning to dominate the mass growth only at $z<1$.  
Note that in hierarchical models, what is true for cluster mass halos is
also true for galaxy mass halos.  While gastrophysics may complicate the
discussion, we raise this as an example where mass growth due to major
mergers does not lead to increased density in the central regions.
This appears to be in remarkable agreement with what we see.  
\textit{If the $z\sim 2$ objects studied by Bezanson et al. (2009) are to 
evolve into the objects in our sample}, then the required mass growth 
is about a factor of 5 (Figure~\ref{fig:comp2}) -- this is larger 
than most minor merger models can accommodate.  It may well be that 
major mergers were required at $z\gsim 1.5$ and that minor mergers become 
the dominant growth mechanism for massive galaxies only at lower redshift 
(Bernardi 2009; Bernardi et al. 2010b). 

Our Figure~\ref{fig:comp1} supports such a picture:  
major mergers would move the $z=2.3$ objects approximately parallel 
to the solid lines in the two panels. A factor of 5 change in mass and 
size would bring them into much better agreement with the dotted and 
dashed $z=0$ relations in the panel on the left, but they would still 
lie slightly above the corresponding line in the panel on the right.  
Subsequent minor mergers would increase the sizes and decrease the 
velocity dispersions, bringing both the sizes {\em and} densities into 
even better agreement. (Note that a small fractional increase in mass 
results in a larger fractional increase in size and an even larger 
fractional decrease in density.  Indeed, because density is proportional 
to $(\sigma/r_e)^2$, minor mergers are a great way to decrease the density 
for a modest change in mass.)  If the high redshift objects indeed have 
$\sigma\sim 400$~km~s$^{-1}$ (as our Figure~\ref{fig:comp1} may suggest), 
the decrease in $\sigma$ associated with minor mergers will 
(in fact, may be required to) bring the number density of large 
$\sigma$ objects into better agreement with that seen locally 
(Sheth et al. 2003).

On the other hand, selection effects (e.g. due to the small volume observed) could limit the detection of these very massive galaxies. For example, 
the sample of ultramassive early-type galaxies of Mancini et al. (2010) is 
selected from a 2 deg$^2$ field. In such a volume, if galaxies of
M$_{\star}\sim 10^{12}$ M$_{\odot}$ are yet present at $z>1.5$, as predicted by 
model like Fan et al. (2010), the number of detections should be around unity 
and therefore not necessary detected. The detection of these massive 
objects (already difficult in the local universe) is thus a challenge 
for the high redshift universe.

\section*{Acknowledgments}
We thank L. Danese, C. Mancini  R. Sheth and P. van Dokkum  for helpful discussions and the 
anonymous referee for comments that have improved the presentation of our results. 
M.B. is grateful for support provided by NASA grant ADP/NNX09AD02G.  
CM and JP acknowledge support from the Marie Curie Excellence Team Grant 
'Unimass', ref. MEXT-CT-2006-042754, of the European Community.
  
Funding for the Sloan Digital Sky Survey (SDSS) and SDSS-II Archive has been
provided by the Alfred P. Sloan Foundation, the Participating Institutions, the
National Science Foundation, the U.S. Department of Energy, the National
Aeronautics and Space Administration, the Japanese Monbukagakusho, and the Max
Planck Society, and the Higher Education Funding Council for England. The
SDSS Web site is http://www.sdss.org/.

The SDSS is managed by the Astrophysical Research Consortium (ARC) for the
Participating Institutions. The Participating Institutions are the American
Museum of Natural History, Astrophysical Institute Potsdam, University of Basel,
University of Cambridge, Case Western Reserve University, The University of
Chicago, Drexel University, Fermilab, the Institute for Advanced Study, the
Japan Participation Group, The Johns Hopkins University, the Joint Institute
for Nuclear Astrophysics, the Kavli Institute for Particle Astrophysics and
Cosmology, the Korean Scientist Group, the Chinese Academy of Sciences (LAMOST),
Los Alamos National Laboratory, the Max-Planck-Institute for Astronomy (MPIA),
the Max-Planck-Institute for Astrophysics (MPA), New Mexico State University,
Ohio State University, University of Pittsburgh, University of Portsmouth,
Princeton University, the United States Naval Observatory, and the University
of Washington.

\begin{figure*}
 \centering
\includegraphics[width=168mm]{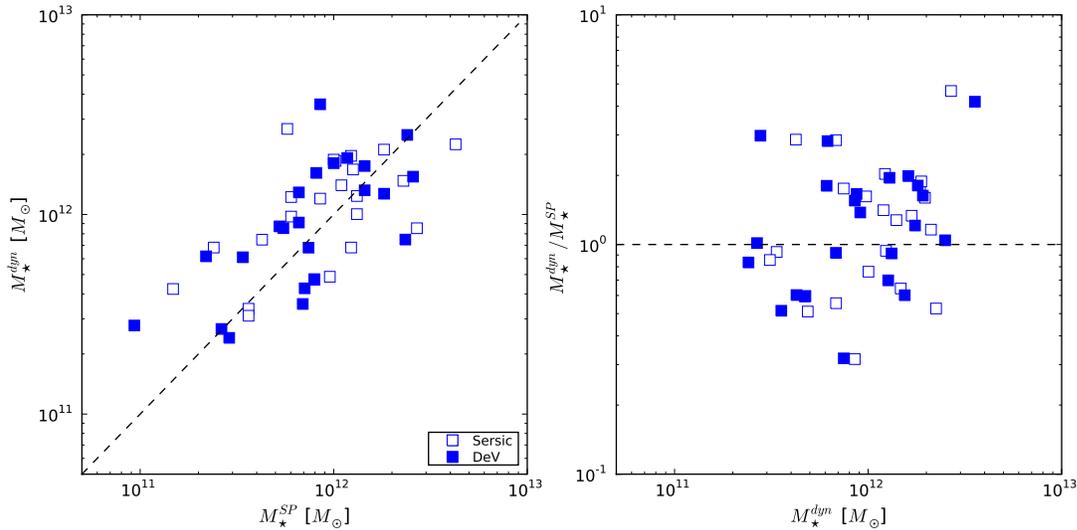}
 \caption{ Same as Figure~\ref{fig:MdynMvis}, but now when $\beta=0.2$ 
          rather than 0.}
  \label{fig:MdynMvisbeta}
 \end{figure*}

{} 

\appendix
\section{Prolate galaxies and anisotropy}

Bernardi et al. (2008) argue that the most massive of these objects 
are likely to be prolate, viewed along the long axis (see alse Thomas et al. 2007).  If the shape 
is due to anisotropic velocities, then the velocity dispersion 
projected along the line of sight is 
\begin{eqnarray}
 \sigma_{los}^2(r) &=&  {{2}\over{\Sigma(r)}} \int_r^\infty {{\rho_\star(R) \sigma_r^2(R)}\over{\sqrt{R^2-r^2}}} R dR \nonumber \\
&&- {{2r^2}\over{\Sigma(r)}} \int_r^\infty {{\beta(R)\rho_\star(R) \sigma_r^2(R)}\over{R\sqrt{R^2-r^2}}}  dR, \nonumber\\
\label{eq:sigma2losbeta}
\end{eqnarray}
(e.g. Binney \& Mamon, 1982), 
where $\beta(r) \equiv 1-\sigma_\theta/(2\sigma_r)$ is the anisotropy 
profile.  When $\beta \rightarrow -\infty $ the orbits are purely 
circular, while  $\beta \rightarrow 1$ corresponds to radial orbits.

We have tried different values of $\beta$ from 0 to 0.5, and found 
that the estimated dynamical mass decreases as $\beta$ increases.  
If $\beta=0.5$ the dynamical estimate of $M_*$ is smaller by 
0.6~dex, but the assumption that all the galaxies of the sample have 
$\beta=0.5$ is inconsistent with most studies which favor $\beta=0$ (e.g. Thomas et al. 2009).  
Figure~\ref{fig:MdynMvisbeta}, which has the same format as 
Figure~\ref{fig:MdynMvis} in the main text, shows results for 
$\beta=0.2$.  The dynamically estimated value of $M_*$ is slightly 
smaller compared to Figure~\ref{fig:MdynMvis}.  

\bsp

\label{lastpage}

\end{document}